# Narratives within Immersive Technologies

Joan Llobera
joan.llobera@starlab.es

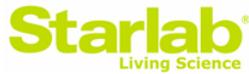

Starlab Barcelona S.L.

Edifici de l'Observatori Camí de l'Observatori s/n

08035 Barcelona, Spain

+34 93 254 03 66

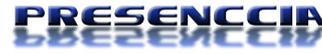

Centre de Realitat Virtual (CRV) Edifici U

Universitat Politècnica de Catalunya

C/Llorens i Artigas 4-6 08028 Barcelona, Spain

## ABSTRACT
The main goal of this project is to research technical advances in order to enhance the possibility to develop narratives within immersive mediated environments. An important part of the research is concerned with the question of how a script can be written, annotated and realized for an immersive context. A first description of the main theoretical framework and the ongoing work and a first script example is provided. This project is part of the program for presence research [16], and it will exploit physiological feedback and Computational Intelligence within virtual reality.

## Categories and Subject Descriptors
H5.1 [**Information, Interfaces and Presentation**]: – Multimedia Information Systems *Artificial, Augmented and Virtual Realities.*

J.5 [Computer Applications]: - Arts and Humanities -*Performance*

## 1. INTRODUCTION
How can we use immersive environments and interactive technologies to develop narratives taking advantage of the presence-enabling phenomena related to them? A specific concern of developing narratives and in general of creative activities is to find language resources specific to a concrete communication media. In this case, the question is how to use virtual immersive environments exploiting the advantages and the limitations the medium provides.

From the point of view of narrative, and even if storytelling is the oldest way of conveying information by humans, all performing arts that derive from it rely on implicit cultural conventions and expectations created by the cultural background of the audience, as well as technical capabilities. The example of the cinema is paradigmatic, and we can trace through history how it gradually separated from theatre and how it generated codes in relation to the technical constraints of the time: from silent cinema to sound cinema, the idea of documentary itself, how post-production techniques influenced audiovisual language, and so on.

Immersive virtual environments provide a new challenge where participants are actually inside and part of unfolding events -even physically impossible events-, participants have in principle complete freedom to attend to any part of the scenario that they prefer, and to go anywhere within it at any time. How is it possible to develop a narrative within this situation?

## 2. NARRATIVES IN IMMERSIVE ENVIRONMENTS
### 2.1 Narratives in theatre
As a first approach, we can look at theatre. By definition, theatre is in 3D, highly immersive, and generally it is limited to audiovisual stimuli. In addition, it generally has an understandable plot, which helps to develop the characters' psychology and depicts a coherent world –different from the real one, but with internal coherence. Even more, something like the *presence* of an actor, which could be understood as "how well he fakes" being the character, is a different but complementary idea of the *presence* or *co-presence* impression we get from an immersive virtual environment.

Performing arts theory has largely described what does a theatre play "work" [11][12]. This last is related with a general hypothesis of reception of the overall process of writing and staging a play.

This process is described as the conscious or unconscious construction of an implicit receptor, related with the expectation/information dialectics: an implicit receptor is supposed to have certain initial expectations that are partially satisfied by certain initial information, but this information itself introduces new expectations, and so on. A reasonable level of satisfaction of the expectations should come with the end of the theatre representation.

The overall process of writing and staging a play implies necessarily a hypothesis of reception, related with the expectations generated by the implicit receptor, the final hypothetic generic spectator, according to the information provided during the play performance.

The way the information is given in traditionally structured plays or in contemporary drama offers a very rich and large set of linguistic resources for a *mise en scène*[*].

## 2.2 Interaction in Presence-enabling environments

Real-time interactions in a virtual environment according to certain rules exist currently in videogame technology, but the way this interaction is implemented is generally quite poor: it generally involves modal interactions like pressing buttons integrated in game pads, which generally don't seem very presence-enabling.

Moreover, this kind of interaction is often incompatible with narrative aspects: the story plot is often an introduction to lead to the interactive part which is the corpus of the videogame itself, or pieces of narration are inserted between pieces of interaction. Even more, it becomes often interaction against narration, being one extreme an ego-shooter like Doom, and the other a movie. From a narrative perspective, the interaction breaks the expectation/information dialectics: the new information consequence of the interaction doesn't change the expectations of the receptor, which in this case is not a *spectator* anymore, but a *participant*.

And what about theatre-like interaction? In drama there are just subtle interactions between public and actors, but there is a lot of interaction between actors. This is also the case in some social scenarios in presence-enabling virtual environments [13][6], but this is not fully automated, as an operator is behind to choose the right answer of the virtual agent.

If we could fully automate those interactions, we could then write and stage scripts in which we have this kind of presence-enabling interactions embedded in an immersive audiovisual environment, with a strong narrative component.

## 2.3 Narratives in Immersive Environments

To deal with the different possible actions of the participant at a certain moment, we are in a framework similar to a cooperative game [8]: we need to suppose a minimum degree of collaboration –like for example to sit down round a table if the collaborative game is a negotiation-, every agent has his own goals, and they all share a common goal, in this case abstract: making the story advance. The general idea is that if the participant doesn't assume a certain action necessary for the plot, a virtual agent can do it instead of him.

A participant appears like a character who can adopt a multitude of roles in a given situation, but a character who doesn't know what role he is expected to adopt, relying only on his empathic abilities and understanding of the situation to interact socially.

---

[*] Merriam-Webster dictionary defines *mise en scène* as *"the arrangement of actors and scenery on a stage for a theatrical production"*.

This kind of character can be exemplified as the Zelig character in the film that has the same name [1].

We chose then a declaration plus instantiation approach. The first is done by a scriptwriter and agreed by a development team. The script is implemented in 2 concurrent processes: an interaction engineering team handles the detection of possible occurrences by estimators, as described in Section 3.

In parallel, a director with the help of actors and/or animators should instantiate all the stated actions in the script, in all the possible configurations according to the interaction variability, as well as the complementary actions he thinks necessary.

The last step would be to adjust the threshold reactivity and the latencies of the action triggers, in a process equivalent to the one of cutting or editing in movie production.

This proposed pipeline presents the advantage compared to the approach introduced in [9], based on Joint Dialog Behaviors, the only fully implemented interactive drama the author knows, that the starting point is a script that doesn't need major technical know-how, and that there is a certain separation of the creative and technical tasks involved in the production process.

As an example of this approach to interactive drama, a short script adapted from an old joke is provided in Example 1, with some writing conventions detailed.

## 3. AUTOMATING NATURAL INTERACTION

### 3.1 Taxonomy of discursive actions

Before we can automate the reaction to an action, we first need to discriminate those actions. The taxonomy of the large set of discursive actions going on at the same time in a theatre play has not been analyzed exhaustively, as far as the author knows. Theatre directors do not explicitly know what the contextual actions that reinforce, deform or invert the meaning of saying a sentence are; they just find the good actions and when to do them as part of the creative process with the actors and the whole *mise en scène*.

However, and even if discursive actions are not analyzed exhaustively in drama theory, linguists have analyzed them quite exhaustively in theatre-like situations, like classes or other natural environments.

This taxonomy of communicational acts involves three complementary research fields: Conversation Analysis [3], Non Verbal Communication [10], and Pragmatics [14]. Roughly, the first explores oral interaction and interaction rules in oral conversations like turn-taking, and has the practical advantage that it can be entirely analyzed from a conversation transcription.

The second explores all the information that is not strictly textual, from gesture to voice intonation through synchronic events on surrounding landscape and epidermis reactions, using large matrices of discourse-relevant events.

The third one explores how meaning of a sentence is completed or modified by the surrounding scene and implicit assumptions done in everyday conversations, like intention attribution.

## 3.2 Feature extraction and classification

Advanced systems for automatic feature extraction and classification can actually perform very sophisticated decisions based on multi-sensor data fusion [5] [4]. automatically a pragmatically relevant subset of the features described in Non-Verbal Communication in order to detect interaction.

For example, we could use verbal intonation, physiologic reaction

**Example 1: a short script for Interactive Drama**

| Conventions: | Script: |
|---|---|
| -Sc stands for "Scene". First paragraph is a general description of the ambient, there are no actions in it.<br>-SS, for "Scene step", thinking of part of a scene, or a state of Finite State Machine.<br>-ZELIG designates the participant.<br>-Stated actions will happen necessarily.<br>-IF *x* THEN *y* are fuzzy conditions\*, which can be true in a variable degree, and more than one at the same time. *x* is a possible linguistic variable, and *y* a necessary action related to the occurrence of *x*.<br>-NOTP is for none of the previous, or not(P), which is the negation of the previous proposition(s). It necessarily appears for logic consistency.<br>-Indentation defines one level or set of possible actions, a subset of which will necessarily happen.\*\*<br>-*next SS* means getting out of a set of IF x THEN y (like a *return* instruction). *continue* means keep going through them<br>-The actions in brackets [] only are executed if they are necessary for logic consistency.<br>-Comments are in parenthesis ()<br>-*or similar* means a detected sentence with similar intentional content, which could be detected with a synonym dictionary (in a large sense) after a speech recognition step. | *Sc1. Exterior Night. A not well illuminated Street.*<br><br>*ZELIG is in a street, quite empty. Under a streetlamp, a drunken guy is looking for something on the floor.*<br><br>*SS1.*<br><br>    IF ZELIG gets surprised of the environment THEN wait.<br><br>    IF NOTP, THEN continue<br><br>*ZELIG approaches the guy, as nothing else is happening (this is the minimum hypothesis of collaboration).*<br><br>*SS2.*<br><br>    IF ZELIG asks him what is the problem, or what's going on, or similar THEN next SS<br><br>    IF ZELIG asks him where are they, or why is he like that, or what's he's name, or any other similar question THEN the drunk guy doesn't really answer, just some drunk guy comments, and continues looking around on the floor.<br><br>    IF NOTP (ZELIG is not proactive enough), THEN a policeman appears and asks the drunk guy what's going on<br><br>*SS3.*<br><br>*The drunken guy gesticulates pointing to the door next to him, and says "I'm looking for my keys".*<br><br>    IF ZELIG starts looking with the drunk guy THEN the policeman [appears, looks what's happening and] looks for the keys.<br><br>    IF NOTP THEN The policeman starts looking with the drunk guy and asks ZELIG to collaborate.<br><br>*SS4*<br><br>*After a while,*<br><br>    IF ZELIG asks to the drunk guy "are you sure of having them lost over here", or similar THEN The drunk guy answers "No, I just lost them over there" and points a dark part, "but there it's to dark to find them, so I look for them here" and next SS.<br><br>    IF ZELIG gets tired of looking and doesn't ask, THEN The policeman asks the question and the drunk guy answers the same.<br><br>    IF NOTP THEN wait until ZELIG gets tired.<br><br>*The streetlamp turns off.*<br><br>END |

\* Very precisely, *IF x THEN y* is equivalent to the Catalan construction *Si x cal y*, and to the English construction *If the x possibility occurs, then y will necessarily occur*, where possible and necessary are fuzzy measures such as nec(a)=1-pos(not(a)), and nec(a) ≤ prob(a) ≤ pos(a)

\*\* Note that the overall structure of SS and possibility/necessity measures in *IF x THEN y* blocks is defined in such a way that it can be recursive: defining SSS would be straightforward, if it was helpful to script a complicated scene.

Given information of oral speech, spatial position, body movement and/or physiologic response, we could extract and gesture to extract emotional content, or extract authority relations according to relative spatial positions, and so on. As well, to trigger the reaction, we would use part of the conversation rules described in Conversation Analysis.

### 3.3 Decision and action

As we automate the discrimination between different communicative actions of the participant(s), the director and his team will stage[†] adequate reactions of the characters.

Note that the interaction is described with sets of *IF x THEN y*, always ended by a *NOTP* possibility. These are fuzzy conditions with a possibility/necessity measure [17], where *x* and *y* are fuzzy linguistic variables [18]. For *x* we suppose, even if it is not a minor assumption, that a probabilistic estimation can be built, which will be embedded in a fuzzy matrix for the decision stage, like the table in Example 3.

As well, for each character other goals will be declared and relevant actions related to it instantiated, in order to help increasing the psychological depth impression we expect from a character when no plot-relevant action needs to be triggered.

Now, let's go back to the collaborative game framework. We suppose that every character (including Zelig) has situation-dependent motivated goals, and they all share the goal "contribute to the advance of the plot" This lets us implement a narrative scripted like above in a distributed way with Extended Behavior Networks [7], by using the necessity measure introduced before as the fuzzy truth value of every scripted proposition. This respects the principle of locality for each character.

### 4. MEASUREMENT

Measurements can be done in order to evaluate 2 aspects. First, we can try to evaluate what we are loosing with the use of immersive technologies comparing the subjective, physiologic and behavioral responses to an (interactive) performance with real and present actors and one with immersive technologies mediation.

A second aspect that must be evaluated is the success of the interaction. As speech recognition, conversation analysis rules, non-verbal interaction and biofeedback will –at least partially- be used for the interaction; it is hard to evaluate its success with those. But we can analyze the pragmatic aspect of the interaction. According to the context and the apparent verbal and non-verbal interaction, we can analyze how well we accomplish principles of relevance, pertinence, etc. and, as well, analyze at what extent participants enter in this pragmatic level of engagement.

For example, if we don't say explicitly a character is angry or in love with someone else, and the participant says "this character was in love with that girl, that's why he was fighting that other character" when this was never said, then we have success, as there is implicit mind attribution, and it will be reflected both in the pragmatic analysis of the scene and in the subjective questionnaire of the participant.

Note that the overall can also be seen as a less ambitious version of the Turing test [15]: we don't try to fake that a computer answers like a human does in any situation. We rather try to prove that, in a certain controlled situation, a computer can fake being another person or a character as well as a person can. A pragmatic analysis of the different characters and the participant interaction should allow us to state at what extent we succeeded.

**Example 2: Fuzzy linguistic Variable**

Let's suppose we have this set of possible actions
    *IF Zelig gets angry THEN Angie gets scared*
    *IF Zelig moves toward the table THEN the drunk man gets in his way*
    *IF NOTP THEN the drunk man gets angry and Angie gets scared*

As "Zelig gets angry" is a linguistic variable, it can be, for example:
"very angry", "not very angry", "very slightly angry" and NOTP.

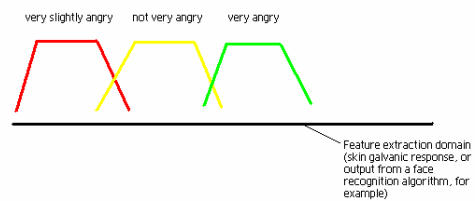

The reaction of Angie can be
A1="cry dramatically", A2="be scared" and A3="stop smiling".
In the same way, the reaction of the drunk guy getting in his way can be B1, B2 and B3, with different levels of intensity.
The drunk man gets angry can be c1.
The work of the interaction engineer would be to extract the features relevant to calculate the different levels of the linguistic variables "Zelig gets angry" and "Zelig moves towards the table", and to define more precisely the crossed interactions of consequences of the possible actions, like it's done in Example 3.
A second step would be to detect possible configurations that would generate action incompatibilities, inform the director and decide with him an appropriate response.
For example, let's imagine that if Angie cries dramatically she also is situated between the table and Zelig, because of the spatial situation of every character.
In this case, the reaction of the drunken guy is not appropriate any more, and this should be stated in the fuzzy action matrix, for example, asking the director to do a new action instantiation, declared in Example 3 as B3.1.

**Example 3: Fuzzy matrix decision**

|  | Z. turns towards the table | Z. walks towards the table | Z. runs towards the table | NOTP |
|---|---|---|---|---|
| Z. is very angry | A1 & B1 | A1 & B2 | A1 & **B3.1** | A1 |
| Z. is not very angry | A2 & B1 | A2 & B2 | A2 & B3 | A2 |
| Z. is slightly angry | A3 & B1 | A3 & B2 | A3 & B3 | A3 |
| NOTP | C1 & B1 | C1 & B2 | C1 & B3 | C1 & A1 |

---

[†] Note that here "to stage" is used as "to plan, organize and carry out" but as well as "to perform (a play)"

## 5. ACKNOWLEDGMENTS

I would like to thanks specially Prof. Llobera for helping me finding the right references for the linguistic aspects.

As well, I acknowledge Prof. Batlle for the drama theory references. As well, thanks to Dr. Ruffini from Starlab and Prof. Slater from Presenccia-UPC, my PhD directors, for their advice and corrections.